\documentstyle[12pt]{article}

\begin{document}

\begin{titlepage}

\vskip 0.9truecm

\begin{center}
{\large {\bf Hydrostatic Equilibrium of a Perfect Fluid Sphere
with Exterior Higher Dimensional Schwarzschild Spacetime}}
\end{center}

\vskip 0.8cm

\begin{center}
{ \sc J. Ponce de Leon}\footnote{E-mail: jponce@upracd.upr.clu.edu or jpdel1@hotmail.com}\vskip 0.2cm
{\it Laboratory of Theoretical Physics, Department of Physics, 
University of Puerto Rico, P.O. Box 23343, San Juan, PR 00931, USA.\\}
\vskip 0.6cm

{ \sc Norman Cruz}\footnote{E-mail: ncruz@lauca.usach.cl}
\vskip 0.2cm
{\it Departamento de F\'{\i}sica,
    Facultad de Ciencia,
    Universidad de Santiago de Chile,
    Casilla 307, Santiago, Chile.\\}

\end{center}

\vskip 0.7cm

\noindent {\small {\bf Abstract:}

We discuss the question of how the number of dimensions of space and time 
can influence the equilibrium configurations of stars. We find that
dimensionality 
does increase the effect of mass but not the contribution of the pressure, 
which is the same in any dimension. In the presence of a (positive) 
cosmological constant the condition of hydrostatic equilibrium imposes a 
lower limit on mass and matter density. We show how this limit depends 
on the number of dimensions and suggest that $\Lambda\,>\,0$ is more effective 
in $4D$ than in higher dimensions. We obtain a general limit for the
 degree of compactification (gravitational potential on the boundary) of 
perfect fluid stars in $D$-dimensions. We argue that the effects of 
gravity are stronger in $4D$ than in any other number of dimensions. 
The generality of the results is also discussed.}
   
\end{titlepage}
\newpage
\section{INTRODUCTION}
Lately the study of higher dimensional space-times has 
led
to important generalizations and wider understanding of general 
relativity
solutions. Of special interest are the black-hole solutions found 
by Myers and Perry \cite{myers}. They 
generalized the Schwarzschild, Reissner-Nordstr\"{o}m 
and Kerr solutions and discussed the associated singularities, 
horizons and 
topologies.
The cosmological constant was included in these solutions by
Xu Dianyan \cite{Xu}. Also, Ba\~{n}ados {\it et al} \cite{banados}, in the
context of Lovelock's theory, 
 found black-hole
solutions where the parity of the dimensions plays an important role.
Interior solutions for higher dimensional perfect-fluid have been 
discussed by Krori {\it et al} \cite{Krori} and Shen {\it et al} 
\cite{Shen}.
Other interior solutions, as well as their 
physical properties,
 have been discussed in the context of ``matter-from-geometry" theory, in which 
four-dimensional matter is 
interpreted as a manifestation of five-dimensional geometry
\cite{{Wesson, JPdL},{JPdL, Wesson}}. 

In this paper we are not interested in obtaining new solutions.
Our object here is to study the general conditions for the hydrostatic 
equilibrium of spherical stars in $D$ dimensions. We 
focus our attention on effects of 
extra dimensions. Conversely, we are interested in what limitations 
observations might put on the number of dimensions. 
We will show that the increase in the number of 
dimensions 
increases the effect of mass. We will see that the degree of 
compactification\footnote{The degree of compactification $GM/R^{D-3}$, 
is a measure of how much mass can be packed in a given volume, 
without provoking gravitational collapse.}, that is the surface gravitational
potential 
of a perfect fluid star, is maximum for $D= 4$. 

The plan of this article is as follows. In Sec. 2 we will discuss
 the effects on   
gravitational mass and pressure in equilibrium configurations.
In Sec. 3 we study the degree of 
compactification of stars in $D$ dimensions, and present an 
explicit example that illustrates our findings. In Sec. 4 we 
summarize our results. 

\section{HYDROSTATIC EQUILIBRIUM IN $D$ DIM}
\subsection{The Tolman-Oppenheimer-Volkov equation}
In this section we will obtain the $D$-dimensional version of the 
{\it Tolman-Oppenheimer-Volkov} equation ($TOV$) and discuss its consequences on
the 
hydrostatic equilibrium of spherical stars in $D$ dim. With this aim, 
and following the conventional wisdom, we assume that the form of the physical 
laws is independent of the number of dimensions. Consequently, we start with 
the Einstein field equations, which in $D$ dimensions are

\begin{equation} 
R_{A B} = 8 \pi G [T_{A B} - {1 \over D-2} g_{A B}
T],
\label{einstein}
\end{equation}
where $G$ is the gravitational constant in $D$ dimensions, and $T_{A B}$ is the 
$D$-dimensional energy momentum tensor (capital indices run over 
1,2,..,D).

Let us consider the $D$ dimensional spherically symmetric metric, given by
\begin{equation}
ds^2 = e^{\nu (r)}dt^2 - e^{\lambda(r)}dr^2 - r^2 d \Omega,
\label{metric}
\end{equation}
where $d \Omega$ is the line element on a unit $(D-2)$ sphere.

Now, let us assume that the $D$ dimensional energy-momentum tensor has
the form
\begin{equation}
T_{A}^{B} = diag(\rho, -p, -p, ..., -p),
\label{momentum}
\end{equation}
where $\rho$ is the energy density and $p$ is the pressure. 
With this choice the field equations (\ref{einstein}) reduce to

\begin{equation}
e^{- \lambda(r)} \left ({\lambda' \over r} - {n \over r^2} \right) + {n \over
r^2} = {16 \pi G \over n+1} \rho,
\label{e-lamda}
\end{equation}

\begin{equation}
e^{- \lambda(r)} \left ({\nu' \over r} + {n \over r^2} \right) - {n \over r^2}
= {16 \pi G \over n+1} p,
\label{e-nu}
\end{equation}

\begin{equation}
e^{- \lambda(r)} \left ({\nu'' \over 2} + {\nu'^2 \over 4} - {\lambda' \nu'
\over 4} - {{n \lambda' + \nu'} \over 2r} - {n \over r^2} \right) + {n \over
r^2} = 0,
\label{e-lanu}
\end{equation}
where $n = D - 3$. In the empty space surrounding the sphere of matter 
the solution to these equations 
is the exterior Schwarzschild solution in $D$ dimensions\footnote{This line
element 
becomes singular on $r = (2GM)^{1/n}$. 
Therefore, in $D$ dimensions, the ``radius" $R$ of a spherical star, in
equilibrium, 
must be larger than $(2GM)^{1/n}$. This is what gives rise to the question of
how close
the ratio (surface gravitational potential) $GM/R^{n}$ can approach the limiting 
value $1/2$.}  
\begin{equation}
ds^{2} = ({1- \frac{2GM}{r^n}})dt^{2} - \frac{dr^{2}}{{1- \frac{2GM}{r^n}}} -
r^{2}d\Omega.
\label{ext Shw}
\end{equation}
where $M$ is interpreted as the total mass of the body. Inside 
the sphere this solution must be continued by a solution to
 (\ref{e-lamda})-(\ref{e-lanu}) obtained under the conditions that
 the pressure  be zero at the
 boundary of the sphere, and the metric functions be continuous across this 
boundary.

Equation (\ref{e-lamda}) is easily integrated into

\begin{equation}
e^{- \lambda} = 1 - {2 G m(r) \over r^n},
\label{sol-lam}
\end{equation}
where the function $m(r)$ is given by

\begin{equation}
m(r)  = {8 \pi \over n + 1} \int_0^r \rho(r') r'^{n + 1} dr'.
\label{mass function}
\end{equation}
and the constant of integration has been set equal to zero 
to remove singularities at the origin.

The continuity of the interior metric with the exterior 
solution at the boundary of the sphere $r = R$, requires that $m(R) =M$. 
Therefore, Eq. (\ref{mass function}) is interpreted as the 
gravitational mass inside a sphere of radius $r$. In what 
follows we will call it
{\it mass function}\footnote{The normalization of this function is different 
from the 
mass function in $D$ dimensions 
as defined in Ref. \cite{myers}, which is
$$
m(r)=A_{n+1}\int_0^r\rho(r')r'^{n+1}dr'
$$
where $A_{n+1}=2\pi^{(n+2)/2}/\Gamma((n+2)/2)$.}.

The conservation of stress
energy yields 
\begin{equation}
p' = -{\nu'\over 2}(\rho+p).
\label{tuv}
\end{equation}
The expression for $\nu'$ can be obtained from equations
(\ref{e-nu}) and (\ref{sol-lam}) 
\begin{equation}
\nu'= {2 G \over n + 1}{m n (n + 1) + 8 \pi p r^{n + 2} \over r (r^n -2 G m)}.
\label{nu}
\end{equation}
Introducing equation (\ref{nu}) into equation (\ref{tuv}) we obtain 
the corresponding {\it Tolman-Oppenheimer-Volkov} equation in $D$ dimensions
\begin{equation}
{dp \over dr} = -{ G (p + \rho) [m n (n + 1) + 8 \pi p r^{n + 2}] \over (n + 1)
r
(r^n - 2 G m)}.
\label{O-V}
\end{equation}
This equation shows how dimensionality affects the equilibrium.

 Firstly, we see that, as in $4D$ general relativity, the pressure acts as a
source
 of the gravitational field (because of the term proportional 
to $p$ added to $m$). Also, it is affected by the 
gravitational field (because $\rho$ is replaced by $(\rho +
p)$). Now, equation (\ref{O-V}) shows that both the contribution of pressure to 
the source and the effects of gravity on 
pressure are independent of the number of dimensions.

 Secondly, we notice 
that the effects
 of mass do increase (decrease) with the increase (decrease) of the number 
of dimensions. This 
follows from the term $n(n + 1)$, which multiplies the gravitational 
mass $m(r)$ in (\ref{O-V}).

Finally, for the sake of completeness, we mention that the term $1 / r^{D -2}$ 
in the Newtonian force is now replaced by $1 / {r^2(r^{D - 4}
- {2 Gm / r})}$.

\subsection{Perfect fluids with cosmological constant}

 The behavior of perfect fluids changes qualitatively when a
cosmological constant is considered. For example, in 2+1 dimensions 
a black hole solution is obtained
with the inclusion of a negative cosmological constant
  \cite{Banados1}.  
Perfect fluid stars, in 2+1 dimensions, should collapse depending
on their degree of compactification \cite{Cruz}.

Now, we will show that in the presence of a (positive) cosmological constant 
the $TOV$ equation (\ref{O-V}) imposes a lower limit on mass and matter density. 

The inclusion of the cosmological constant,
$\Lambda$, in the Einstein equations is straightforward making the replacements
$\rho\rightarrow \rho + \Lambda$ and $p\rightarrow p - \Lambda$.
Eq. (\ref{sol-lam}) becomes

\begin{equation}
e^{- \lambda} = 1 - {2 G\mu(r) \over r^n}.
\label{sol-con}
\end{equation}
where $\mu(r)$ is given by

\begin{equation}
\mu(r) = m(r) + {8 \pi \Lambda r^{n+2} \over (n+1)(n+2)}.
\label{mass with Lambda}
\end{equation}

If we evaluate the $TOV$ equation on the surface of the fluid
$(r=R)$ we obtain

\begin{equation}
p'(R)= {-G \rho \left[n(n+1)M - 16 \pi \Lambda R^{n+2}/(n+2) \right] \over
(n+1) R (R^n -2G\mu(R))}.
\label{plam}
\end{equation}

Hydrostatic equilibrium requires $p'\leq {0}$. Therefore, 
the right hand side of
equation (\ref{plam}) implies a lower bound on $M$, namely

\begin{equation}
M \geq {16 \pi \Lambda R^{n+2} \over n(n+1)(n+2)}.
\label{Mbound}
\end{equation}

This bound does not depend on the sign of $\Lambda$. It is, of
course, significant when $\Lambda\,>\,0$.

The above equation implies a lower limit on the matter 
density. In the particular case of fluid with uniform density,
$\rho = \rho_0$, this lower bound is given by
\begin{equation}
\rho_0 \geq {2 \Lambda \over n}.
\label{roconst}
\end{equation}

We note that $\rho_0$ decreases with the 
increase of the number of dimensions, which indicates that 
a positive cosmological constant is less effective producing 
repulsion in higher dimensions than in $4D$. In other words, in $4D$ 
the repulsion produced by $\Lambda\,>\,0$ is bigger and consequently 
more mass is needed to balance the distribution.

\section{DEGREE OF COMPACTIFICATION OF STARS IN $D$ DIMENSIONS}

\subsection{Extension of Buchdahl's Theorem to $D$ dim}

In general relativity it is well known that the equations of
stellar structure for perfect fluid matter lead to the
existence of an upper mass limit, $viz.$,

\begin{equation}
{GM \over R} \leq {4 \over 9}.
\label{MR}
\end{equation}

This result, which was first showed by Buchdahl \cite{Buchdahl}, 
is valid under very general conditions irrespective of the equation of
state. These conditions are: (i) The material of the sphere 
is locally isotropic. (ii) The energy
density is positive and does not increase outward. The degree of
compactification 
in spherical stars is fixed by the contribution from the Weyl curvature tensor 
(the ``purely gravitational field energy") to the mass-energy inside
 the body  \cite {{Ponce de Leon 10},{Ponce de Leon 11}}.

In this section we show how 
Buchdahl's result (\ref{MR}) can be extended to any 
number of dimensions under the same physical conditions.

 First we study 
the isotropy condition given by Eq. (\ref{e-lanu}). This condition has 
extensively been studied in $4D$, where a considerable simplification 
is attained with the introduction of the following notation \cite {Ponce de Leon
12}

\begin{equation}
 e^{- \lambda} = 1 - {2 Gm(r) \over r^n} = Z, \;\; 
e^{\nu(r)} = Y^{2}, \;\;  r^2 = x.
\label{appropiate notation}
\end{equation}
and 
\begin{equation}
u =  \int^x_0 {d x'\over\sqrt{Z(x')}}\;.
\label{u variable}
\end{equation}
With this notation Eq. (\ref{e-lanu}) reduces to 
\begin{equation}
2{d^{2}{Y}\over du^{2}} = nYG{d \over dx} \left ({m \over r^{n+2}} \right )\;.
\label{new isotropy}
\end{equation}

The term $m/r^{n+2}$ can be identified with the mean density of
the fluid sphere in $D$ dimensions. Therefore, our second condition 

\begin{equation}
{d \rho \over dr} \leq 0,
\label{condition 2}
\end{equation}
requires

\begin{equation}
{d \over dr} \left ({m \over r^{n+2}} \right ) \leq 0.
\label{average density}
\end{equation} 

Now Eq. (\ref{new isotropy}) gives

\begin{equation}
{d^{2} Y \over du ^2} \leq 0,
\label{second derivative}
\end{equation}
which means that ${dY/du}$ decreases monotonically. This in turn implies   
\begin{equation}
{d Y \over du } \leq {{Y(u) - Y(0)} \over u}.
\label{mean value theorem}
\end{equation}

Since both $Y(0)$ and $u$ are non-negative, it follows that

\begin{equation}
Y^{-1} {dY \over d u } \leq \frac{1}{u}.
\label{first derivative of Y}
\end{equation}
In term of the original variables this equation reads

\begin{equation}
\left (1 - {2 Gm(r) \over r^n} \right)^{1/2} {d \nu \over dr}
\leq r \left [ \int^r_0 d r'
r' \left (1 - {2 Gm(r') \over r'^n} \right)^{-1/2}  \right ] ^{-1}.
\label{basic inequality}
\end{equation}

Using the fact that the average density decreases outward, we can evaluate 
the integral in (\ref{basic inequality}) as follows 

\begin{eqnarray}
\int^r_o d r' r' (1 - {2 Gm(r') \over r'^n})^{-1/2} \geq \int^r_o
dr' r' \left  (1 - {2G m(r) \over r^{n+2}} r'^2 \right )^{-1/2} \nonumber \\
= {r^{n+2} \over 2 Gm(r)} \left [ 1- (1-{2 Gm(r) \over r^n})^{1/2} \right ].
\label{evaluation}
\end{eqnarray}
Now, substituting (\ref{nu}) and (\ref{evaluation}) into (\ref{basic
inequality}) 
we obtain

\begin{eqnarray}
\frac{{n(n+1)}m(r)/r^n + 8\pi Gpr^2}{(n+1) (1-{2G m(r) \over
r^n})^{1/2}} \leq {2 G m(r) \over r^n} \left
[ 1-(1- {2 Gm(r) \over r^n})^{1/2} \right ]^{-1}.
\label{Buchdahl}
\end{eqnarray}
This equation can be simplified as follows

\begin{equation}
\frac{Gm(r)}{r^{D - 3}} \leq \frac{1}{(D -1)^2} \left [(D - 2) - \frac{D - 1}{D
- 2}8 {\pi} G pr^2 +  \sqrt {(D - 2)^2 + \frac{D - 1}{D - 2}16{\pi} Gpr^2}
\right].
\label{degree of compactness}
\end{equation}

Finally, the degree of compactification is obtained by evaluating this
expression 
at the surface of the fluid $r = R$, where $p = 0$. We obtain 

\begin{equation}
\frac{GM}{R^{D - 3}} \leq \frac{2(D - 2)}{(D - 1)^2}.
\label{D Buchdahl}
\end{equation}
This is the desired extension 
to $D$ dimensions of Buchdahl's theorem. Indeed for $D = 4$ we recover
Buchdahl's 
limit.  It shows that $GM/R^{D-3}$ decreases with the increase of the number
of dimensions. Consequently, the degree of
compactification is maximum in $D = 4$.

\subsection{Particular example}

The results discussed so far  are model independent; they follow
solely from the 
physical conditions (1) and (2) listed in Sec. 3.1, and the matching conditions
at the boundary surface. In this section we show a simple model, with
``reasonable" 
physical properties,  
that illustrates our results and indicates how (in what limit) our inequality
(\ref{D Buchdahl}) 
can be saturated.

We will consider the case of uniform proper density in $D$ dimensions. With 
$\rho = \rho_0$, Eqs.(\ref{mass function}) and (\ref{u variable}) give
\begin{equation}
{m(r)} = \frac{8{\pi} {\rho_0} r^{n+2}}{(n+1)(n+2)},
\label{massa Shw}
\end{equation}
and 
\begin{equation}
u = {{2}\over{C}}\left[1 - \sqrt{1 - Cx}\right],\;\;\;{C}= \frac{16{\pi}G
\rho_0}{(n+1)(n+2)},
\label{u Shw}
\end{equation}
respectively. Now, the isotropy condition (\ref{new isotropy}) becomes $Y_{uu}$
= $0$.

Therefore
\begin{equation}
Y = e^{\nu/2}= A +B\left[1 - \sqrt{1 - Cx}]\right.
\label{Y Shw}
\end{equation}
The constants $A$, $B$ and $C$ are specified by matching the interior metric 
to the exterior Schwarzschild solution (\ref{ext Shw}). 
The result is 
\begin{equation}
Y = e^{\nu/2} = {(1 +\frac{n}{2})}\sqrt{1- \frac{2GM}{R^n}} -
 {{n}\over{2}}\sqrt{1- \frac{2GMr^2}{R^{n+2}}},
\label{int metric g zero zero}
\end{equation}

\begin{equation}
Z = e^{-\lambda} = {1- \frac{2GMr^2}{R^{n+2}}}.
\label{interior metric g one one}
\end{equation}

For the pressure and density we find
\begin{equation}
{{p}\over{\rho}} = {n}\;\;\frac{\sqrt{1- \frac{2GMr^2}{R^{n+2}}}
- \sqrt{1- \frac{2GM}{R^n}}}{(2 + n ) \sqrt{1- \frac{2GM}{R^n}} - n\sqrt{1-
\frac{2GMr^2}{R^{n+2}}}}.
\label{pressure Shw}
\end{equation}

Notice that $p = 0$ at the boundary $r = R$ and $(p/\rho) \geq 0$. In order that 
the pressure may never become infinite at the origin the denominator of 
(\ref{pressure Shw}) must never vanish. This gives

\begin{equation}
\frac{GM}{R^n}\ < \frac{2(1+n)}{(2 + n)^2} = \frac{2(D -2)}{(D - 1)^2},
\label{mass limit Shw}
\end{equation}
which is the limit (\ref{D Buchdahl}) found 
from general considerations. We notice that the 
equality

\begin{equation}
\frac{GM}{R^n}\ = \frac{2(1+n)}{(2 + n)^2} = \frac{2(D -2)}{(D - 1)^2},
\label{saturation}
\end{equation}
takes place in the limit when the central pressure 
becomes infinite. This is exactly what happens with the famous $4/9$ Buchdahl 
limit in $4-D$.

\section{CONCLUSIONS}

Equation (\ref{D Buchdahl}) constitutes our main result. It 
shows how the degree of compactification depends on the 
number of dimensions. At the surface of the star the  metric 
coefficient $g_{00}$ is given by
$e^{\nu (R)} \geq {(D -3)^2}/{(D - 1)^2}$. We see that, as a 
function of $D$, the minimum value of $e^{\nu (R)}$ is attained 
for $D= 4$, {\it viz.}, namely $e^{\nu (R)} = 1/9$. For bigger 
number of dimensions $g_{00}(R)$ approaches unity. Since the strength 
of gravity depends on the deviations 
of the metric from its galilean values, it follows that, in more than 
4 dimensions, the classical relativistic effects  are weaker than in $4D$.

This result is obviously related to the fact, elucidated by eq. (\ref{O-V}),
that the effect of 
mass is increased in more that $4$-dim., while the effects of 
pressure remain the same. Consequently, in more than 
$4D$ we cannot pack ``too much" mass into a fixed volume 
without provoking gravitational collapse. In this sense $4D$ is the 
optimum number of dimensions for gravity.

The questions of why does this upper limit (\ref{D Buchdahl}) exist, 
and how it can be saturated, removed 
or at least increased are interesting questions. Previous investigations in $4D$ 
show that the behavior of the Weyl curvature tensor plays 
an important role in this problem \cite{Ponce de Leon 10}, \cite{Ponce de Leon
12}. 
Namely, that a positive (negative) contribution from the Weyl tensor tends to 
increase (decrease) the effective gravitational mass. Furthermore, it has been
shown 
that 
Buchdahl's limit may be increased (or removed) only in the case when 
this contribution is negative. 

Examples of reasonable physical systems showing these properties,  
are provided by certain types of anisotropic fluids. One would expect that 
similar results can be obtained in more than four dimensions.  

  The existence of the lower mass limit, indicated by eq. (\ref{Mbound}),
in the presence of a positive 
cosmological constant, is not surprising. Indeed, the repulsion 
produced by $\Lambda\,>\,0$ should produce expansion (inflation) of an 
empty space. Therefore, some 
minimum amount of mass producing gravitational attraction 
and balancing the repulsion
is needed for an equilibrium to exist. What is interesting here, is the role
played 
by dimensionality. Specifically, that the minimum amount of mass 
depends on the number of 
dimensions and it is maximum in $4D$ and 
decreases with the increase of dimensionality. What this suggests is that 
the effects of a positive cosmological constant would be more important in 
$4D$ than in any other number of dimensions. This again, is a consequence 
of the fact that the effects of mass are increased by dimensionality.



\ ACKNOWLEDGMENTS

The work of N. Cruz was partially supported by Grant 195.0278 from
FONDECYT (Chile) and grant from DICYT of
Universidad de Santiago N$^0$ 0497-31 CM.

\end{document}